\documentclass[aps,prb,reprint,amsmath,amssymb,superscriptaddress]{revtex4-1}

\usepackage{graphicx} 
\usepackage{tikz}  
\usepackage{amsmath}    
\usepackage{amssymb}  
\usepackage{bm}
\usepackage{float}   
\usepackage[colorlinks,citecolor=blue,linkcolor=blue,urlcolor=blue,plainpages=false,pdfpagelabels]{hyperref}      
\usepackage{yhmath}
\usepackage{subcaption}
\begin{document}

\title{Resonant spin Hall and Nernst effect in a nanoribbon of a spin-orbit coupled electronic system}
\author{Mohamad Usman}
\affiliation{Department of Physics, Jamia Millia Islamia, New Delhi-110025, INDIA}
\author{Tarun Kanti Ghosh}
\affiliation{Department of Physics, Indian Institute of Technology-Kanpur, Kanpur-208016, INDIA}
\author{SK Firoz Islam}
\thanks{Corresponding author: s\_islam2@jmi.ac.in}
\affiliation{Department of Physics, Jamia Millia Islamia, New Delhi-110025, INDIA}


\begin{abstract}
We present a theoretical study of spin Hall phenomenon in a nanoribbon of a two-dimensional electronic system with Rashba and Dresselhaus spin-orbit coupling.
We model the electronic system by a square lattice in real space. We show that such nanoribbon can give rise to a number of additional spin degeneracy points as well as anticrossing points, apart from the $\Gamma$ point, between two opposite spin subbands.  We compute the SHC and demonstrate that it diverges and gives rise to a resonance when the chemical potential passes through those spin degenerate or anticrossing points. Contrary to the previous studies, here such resonance emerges even without any external perturbation like magnetic field or light. We also examine the spin Nernst effect and find that it shows clear peaks at the anticrossing and spin degeneracy points, consistent with the Mott relation at low temperature.  Finally, we also investigate the signature of such additional spin degeneracy and anticrossing points in the longitudinal conductance by using the retarded Green function approach in lattice model. The finite width induced subbands are reflected in the longitudinal conductance, 
which takes quantized values of $2n e^{2}/{h}$ where $n$ denotes the number of bands occupied by the chemical potential with each band having spin split subbands. We also note that anticrossing that occurs at low energy between two opposite spin subbands could be also detected via longitudinal conductance.

\end{abstract}


\maketitle

\section{Introduction}
The spin Hall effect (SHE)~\cite{PhysRevLett.83.1834,PhysRevLett126603,
RevModPhys.87.1213,murakami2003} is the hall mark of spin-orbit coupling (SOC) term in a two-dimensional ($2$D) fermionic system in which fermions with opposite spins are scattered to opposite transverse edge normal to the applied bias. The spin-orbit interaction is the result of relativistic effect, and in a $2$D electronic system it mainly arises due to inversion symmetry breaking field across the semiconductor heterojunctions~\cite{Winkler,RevModPhys.76.323}. There are two types of spin-orbit interactions, namely Rashba spin-orbit interaction (RSOI) and Dresselhaus spin-orbit interaction (DSOI). The RSOI arises due to the lack of structural inversion symmetry across the quantum well in a semiconductor heterojunction~\cite{rashba1984} , whereas the DSOI is the result of bulk inversion symmetry breaking~\citep{dresselhaus1955}. The RSOI is externally tunable by applying a gate voltage~\cite{PhysRevB.55.R1958,PhysRevLett.78.1335,PhysRevB.57.11911}, whereas the DSOI is an intrinsic property of the system that cannot be enhanced or tuned through external perturbations. The spin-orbit interaction removes the spin degeneracy even in the absence of a magnetic field. Typical materials that exhibit RSOI include indium-based compounds such as InAs and heterostructures like GaInAs/GaAlAs~\cite{PhysRevB.94.035444,PhysRevLett.78.1335,PhysRevB.55.R1958,PhysRevB.61.15588}. The degree of spin separation along the transverse direction to the applied bias is generally quantified by spin Hall conductivity that was predicted to exhibit a universal constant value in 2D electron gas ($2$DEG) with $k$-linear RSOI~\cite{PhysRevLett126603}.  Subsequently, the spin Hall conductivity in a $2$D heavy-hole gas ($2$DHG) with $k$-cubic RSOI was also carried out and found that instead of universal constant value it's sensitive to the strength of RSOI~\cite{PhysRevB.69.241202,PhysRevB.71.085308}. Several experimental works also reported SHE phenomenon in semiconductor heterostructure ~\cite{kato,PhysRevLett.94.047204,Valenzuela2006,PhysRevLett.98.156601}.

Apart from the above mentioned works on SHE, usual integer quantum Hall phenomenon in such spin-orbit coupled system were also carried out  
~\cite{PhysRevB.41.8278,PhysRevB.67.085313,Mawrie_2014,Mawrie_2017,PhysRevB.98.155442}. In fact, the presence of both the RSOI and DSOI have been also considered in several theoretical works~\cite{PhysRevB.73.155328,VASILOPOULOS2006359,PhysRevB.72.165335,PhysRevB.72.085344,PhysRevB.78.035336,PhysRevB.105.245405}. A series of theoretical investigations on spin-related transport phenomena in $2$DHG in presence of weak magnetic field have also been studied~\cite{PhysRevLett.121.087701,
PhysRevB.101.121302,PhysRevLett.121.077701,cullen.2020}. Moreover, the search for the fermionic system or mechanisms to generate giant or large SHC~\cite{Zhueaav8025,PhysRevB.101.064430,PhysRevB.101.094435} continues to be an active topic in spintronics. 

It is to be noted that the SHE is also one of the key signatures of spin-Hall edge modes in the quantum-spin-Hall liquid discovered in 2D Dirac materials with spin-orbit coupling~\cite{PhysRevLett.95.226801,PhysRevLett.109.055502}. The key difference with quantum spin-Hall insulator is that it belongs to $\mathbb{Z}_2$ topological insulator where bulk is insulating gapped with counter propagating spin-dependent gapless edge modes, whereas semiconductor heterostructure is not topological insulator. 
 
While the SHE describes the generation of transverse spin currents driven by an electric field, its thermal counterpart, The spin Nernst effect (SNE) arises when a longitudinal temperature gradient produces a transverse spin current through the same underlying SOC induced deflection mechanisms. In analogy with the spin Hall conductivity, the strength of this thermally driven response is characterized by the spin Nernst coefficient (SNC), which captures how efficiently a material converts thermal bias into transverse spin flow.
Earlier works in mesoscopic crossbar with RSOI has demonstrated that a thermal gradient generates transverse spin current, and the spin Nernst response exhibits strong dependence on Fermi level, disorder, and external Magnetic field. In magnetic field driven systems, the Nernst type response was found to display sharp enhancements whenever the chemical potential crosses Landau levels~\cite{PhysRevB.78.045302}. In graphene nanoribbon geometries, it has been found that the Nernst response strongly depends on ribbon width and edge chirality~\cite{PhysRevB.80.235411}. More recently, first principles Berry phase calculations in noncollinear antiferromagnets have predicted very large SNC values~\cite{PhysRevB.96.224415}.
Additionally, considerable effort is being devoted to an electronic system with RSOI in confined geometries to explore the interplay of quantum confinement and spin-degree of freedom in transport properties~\cite{zpvy-t4d4, 6wnf-b5g8}. Hence electronic systems with RSOI continues to be an active research field, particularly in the context of spin dependent quantum phenomena~\cite{PhysRevB.111.155411, ty86-f2yt}.

The present work is motivated by the phenomenon called spin Hall resonance based on spin-orbit coupled quantum Hall system, predicted by Shen \textit{et al.}~\cite{PhysRevLett.92.256603}, in which the spin Hall conductance (SHC) exhibits a resonance if spin splitting vanishes between two nearest Landau levels with opposite spin at chemical potential. This study  was also extended to a $2$DEG with the $k^3$-RSOI~\cite{doi:10.1063/1.2345024}, and in a $2$DEG with the presence of both, the RSOI and DSOI~\cite{10.1063/1.2936936}. Recently, one of us has shown that even without Landau level such resonance can occur by applying linearly polarized light, in a $k$-cubic Rashba system ~\cite{k3resonance}. We also note that anticrossings between nearest subbands with opposite spin polarization, accompanied by peaks in the SHC, were previously reported in electronic systems with RSOI in the presence of an additional fictitious lateral parabolic confinement potential~\cite{PhysRevB.60.14272}. However, a parabolic confinement potential is also the result of perpendicular magnetic field. Hence, these two works conceptually resemble to each other, except in presence of magnetic field additional splitting (Zeeman splititng) competes with RSOI and yields a complete vanishing of spin splitting and resulting resonance in the SHC~\cite{PhysRevLett.92.256603}. Whereas, in presence of a fictitious confinement potential there is no Zeeman term and resonance does not occur, rather several anticrossing appears.  We show that such resonance can occur between two subbands with opposite spin in a finite-size system where the width is relatively much smaller compared to length. In fact, we note that a large number of spin splitting vanishing points and anticrossing points can emerge at different energy ranges among the higher subbands. In this work, we discretize the system on a square lattice and consider both normal edge and zigzag edge nanoribbons. We analyze their band spectrum and subsequently study the spin Hall conductivity and spin Nernst effect. We show that for a finite width with RSOI and DSOI, a number of crossing and anticrossing occurs between two subbands with opposite spins, yielding resonance in SHC and Nernst effect.

 This paper is organized as follows. In Sec.~\ref{sec:Model Hamiltonian and Discretization}, we introduce the model Hamiltonian and its discretization. Section~\ref{sec:Energy Spectrum of the Nanoribbon} is devoted to the band structure of the nanoribbon. The Transport properties along with the effects of anisotropy and finite temperature on the SHC are presented and discussed in Sec.~\ref{sec:Transport Properties of the nanoribbon}. and finally, Sec.~\ref{sec:summary} summarizes our findings and provides concluding remarks.


\section{Model Hamiltonian and Discretization}
\label{sec:Model Hamiltonian and Discretization}
In this section,  we briefly discuss the electronic system first. Our objective is to study the spin transport with the finite size effects in a $2$D electronic system. In order to include these effects, we need to work in a real space basis by discretizing the well-known spin-orbit coupled low energy effective Hamiltonian~\cite{rashba1984,dresselhaus1955} 
\begin{equation}
H=-\frac{\hbar^2 \nabla^2}{2m^{\ast}}
-i\alpha\left(\sigma_x \partial_y - \sigma_y \partial_x \right)
-i\beta \left(\sigma_x \partial_x-\sigma_y \partial_y  \right),
\label{eq:cont_ham}
\end{equation}
where $\partial_j = \partial/\partial j$ ($j=x,y$), $\alpha$ and $\beta$ are the strength 
of RSOI and DSOI respectively, $\boldsymbol{ \sigma} =\{\sigma_x,\sigma_y\}$ are Pauli matrices in real spin basis, and $m^{\ast}$ is the effective mass of the electron. A number of works have already considered lattice model of the above Hamiltonian~\cite{PhysRevB.72.165335,PhysRevB.110.045204}. However, we briefly outline the discretization of the above mentioned Hamiltonian by considering a square lattice with lattice constant a along both the $x-$ and 
$y-$ directions. We use the plane wave representation of annihilation operator in the basis of lattice sites \((j,l)\), $c_{\bm{k},\sigma} \;=\; \frac{4\pi^2}{a^2}\sum_{j,l} e^{-i a (k_x j + k_y l)}\, c_{j,l,\sigma}$, we obtain the substitution rules for the continuum momentum operators (with \(\lambda=x,y\))
\begin{equation}
k_\lambda \mapsto \frac{1}{a}\int_{\mathrm{BZ}} d^2k \sin(a k_\lambda) c^\dagger_{\bm{k},\sigma}c_{\bm{k},\sigma}
\label{eq:k_mu}
\end{equation}
\begin{equation}
k_\lambda^2 \mapsto \frac{2}{a^2}\int_{\mathrm{BZ}} d^2k  [1-\cos(a k_\lambda)]c^\dagger_{\bm{k},\sigma}c_{\bm{k},\sigma}
\label{eq:k_mu2}
\end{equation}
using the Eq.~\ref{eq:k_mu2} in the kinetic energy operator 
\begin{equation}
H_{kin} \;=\; \frac{\hbar^2}{2m^*}\int_{\mathrm{BZ}} d^2k\; \frac{2}{a^2}\big[2-\cos(a k_x)-\cos(a k_y)\big]\, c^\dagger_{\bm{k},\sigma}c_{\bm{k},\sigma}
\end{equation}
Performing the Fourier transform to real space yields
\begin{equation}
H_{kin}= -t\sum_{j,l,\sigma,s}
\Big(
c^\dagger_{j,l,\sigma}c_{j+s,l,\sigma}+c^\dagger_{j,l,\sigma}c_{j,l+s,\sigma}\Big)+4t\sum_{j,l,\sigma} n_{j,l,\sigma}
\end{equation}
where $t = \hbar^2/(2m^\ast a^2)$, $ \sigma: \uparrow, \downarrow $, $s= \pm 1$ 
and $ n_{j,l,\sigma} = c^\dagger_{j,l,\sigma}c_{j,l,\sigma}$ .
The on-site term $4t\sum_{j,l,\sigma} n_{j,l,\sigma}$ is an overall energy shift and can be omitted by redefining the zero of energy, yielding the standard nearest-neighbor tight-binding kinetic term:
\begin{equation}
H_{kin}= -t\sum_{j,l,\sigma,s}
\Big(
c^\dagger_{j,l,\sigma}c_{j+s,l,\sigma}+c^\dagger_{j,l,\sigma}c_{j,l+s,\sigma}\Big)
\label{eq:Hkin_lattice}
\end{equation}

In the similar fashion, the RSOI and DSOI terms can also be represented in discrete lattice space. 
Using Eq.~\eqref{eq:k_mu} and
Eq.~\eqref{eq:cont_ham}, we obtain the second quantized form 
of the RSOI~\cite{rashbatb} 
and DSOI~\cite{PhysRevB.72.165335} as

\begin{equation}
H_R =- \frac{i \alpha}{2a} \sum_{j,l,\sigma,\sigma',s } s
\left( \sigma_x^{\sigma\sigma'} c_{j,l\sigma}^\dagger c_{j,l+s,\sigma'}
- \sigma_y^{\sigma\sigma'} c_{j,l,\sigma}^\dagger c_{j+s,l,\sigma'} \right),
\label{eq:HR_final}
\end{equation}
\begin{equation}
H_D = -\frac{i\beta}{2a} \sum_{j,l,\sigma,\sigma',s } s
\left( \sigma_x^{\sigma\sigma'} c_{j,l,\sigma}^\dagger c_{j+s,l,\sigma'}
- \sigma_y^{\sigma\sigma'} c_{j,l,\sigma}^\dagger c_{j,l+s,\sigma'} \right),
\label{eq:HD_final}
\end{equation}

The full tight-binding Hamiltonian including kinetic energy and both the SOC contributions is thus
\begin{equation}
H = H_{kin} + H_R + H_D .
\end{equation}

\section{Energy Spectrum of the Nanoribbon}
\label{sec:Energy Spectrum of the Nanoribbon}
\begin{figure}
    \centering
    \includegraphics[width=0.95\linewidth]{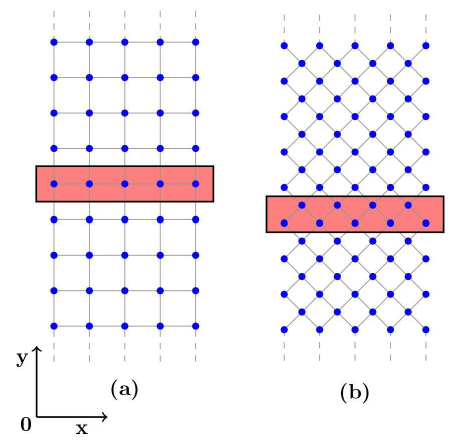}
    \caption{Schematic plots of (a) straight edge nanoribbon of a square lattice, and 
(b) zigzag edge nanoribbon of square lattice. The unit cell is shown 
by the rectangular region highlighted by red colour.}
    \label{fig:Lattice}
\end{figure}

Having established the tight-binding Hamiltonian in real space and its discretization on a square lattice, we now turn to the nanoribbon geometry, which is the central focus of this work. In particular, we analyze two types of nanoribbons, namely straight edge and zigzag edge geometries as shown in Fig.~\ref{fig:Lattice}(a) and Fig.~\ref{fig:Lattice}(b), respectively. To obtain the energy spectrum for nanoribbons with both types of edge terminations, we construct a Hamiltonian analogous to the infinite lattice case~\cite{nanoenergy1,nanoenergy2}. We define a nanoribbon as a stripe of finite width along the $x$-direction whereas infinitely extended along the $y$-direction. Because of the translational symmetry and lattice periodicity along the $y$-direction, we can write down the  total Hamiltonian by using Bloch's theorem as 
\begin{equation}
H = H_{00} + H_{10} e^{ik_ya} + H_{-10} e^{-ik_ya},
\label{ham_eq}
\end{equation}
where $H_{00}$ is the  on-site hopping matrix including the spin-orbit coupling of each supercell as highlighted in red colour in Fig.~\ref{fig:Lattice}, and 
$H_{10}$ denotes the hoping matrix between two nearest supercells. 
The width of the ribbon is described by the number of atomic sites in each supercell. Note that because of the translational symmetry along the $y$-direction, $k_y$ is a conserved quantity and acts as a good quantum number. The energy eigen values are computed numerically by diagonalizing Eq.~(\ref{ham_eq}).
The tight-binding Hamiltonian with spin degrees of freedom gives rise to $N$ transverse bands. In the presence of RSOI and DSOI, each band is split into two in-plane spin-polarized subbands, leading to a total of $2N$ energy subbands. By denoting each subband by $n$,  an energy eigen ket can be fully described by $|\zeta\rangle=\{s, k_y, n\}$ where $s=+(-)$ denotes spin-up (down) subband, respectively.

\begin{figure}
\includegraphics[height=6.6cm,keepaspectratio]{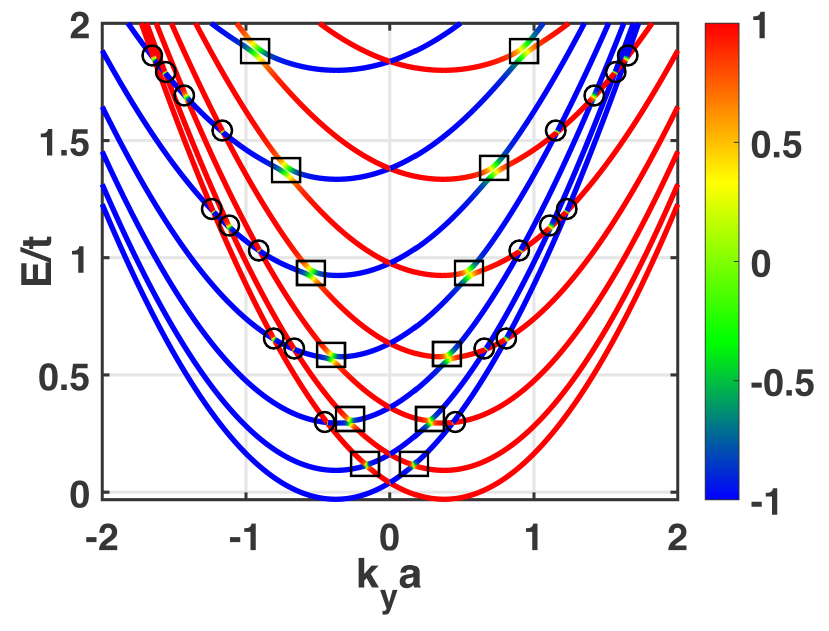} 
\caption{
Spin resolved band dispersion is plotted for a zigzag edge nanoribbon with
$\alpha=0.1$ and $\beta=0.09$ in units of $ta$.
The colour represents the expectation value of the in-plane spin operator,
i.e., the spin polarization in units of $\hbar/2$.
Anticrossing points are highlighted by squares, while spin degeneracy points are
marked by circles.
} 
\label{zigzag_shc_band}
\end{figure}

We plot spin resolved energy spectrum for zigzag nanoribbon around $\Gamma$ point in Fig.~\ref{zigzag_shc_band} for lowest few subbands. 
 The in-plane spin polarization is defined as $\langle \mathbf{S}  \rangle= (\hbar/2) \langle\zeta|\boldsymbol{\sigma}|\zeta\rangle$. In this work, the terms spin up and spin down refer to states with opposite projections of $\langle \boldsymbol{\sigma}  \rangle$ which are distinguished by different colours in the band structure. The energy spectrum is referenced from a baseline value of $4t$, which corresponds to a constant energy shift applied to the full Hamiltonian. The straight edge band spectrum is given in the Appendix \ref{app:straight_shc_band}, as it's qualitative features are essentially the same for lowest few bands.

In the bulk system, spin degeneracy is guaranteed at the $\Gamma$ point by time reversal symmetry (TRS). This characteristic is preserved in the nanoribbon geometry because the confinement does not break TRS. Therefore, TRS imposes Kramers degeneracy $E_{n,k_y,\uparrow}=E_{n,-k_y,\downarrow}$  so that all subbands remain spin degenerate at $k_y=0$. In addition, several extra spin degeneracy points, at which the spin splitting
vanishes, emerge between different subbands with opposite spin. Furthermore, multiple anticrossing points between subbands with opposite spin polarization appear as well. It is important to emphasize that anticrossing points are distinct from spin degeneracy points. At a spin degeneracy point the spin splitting vanishes, yielding a degeneracy of the same order as that observed at $k_y=0$. In contrast, at an anticrossing point the two bands approach each other closely but remain separated by a finite energy gap, indicating an avoided crossing with a clearly resolvable splitting.

We attribute these additional degeneracies and anticrossings to the competition between the RSOI and DSOI i.e., only RSOI or DSOI alone cannot give rise to such degeneracy points and anticrossing points. This is very much similar to the case of Rashba spin-orbit coupled quantum Hall system~\cite{PhysRevLett.92.256603}, where such additional spin degeneracy is the result of the competition between the Rashba and Zeeman splitting. However, in the present case the Zeeman splitting is absent and its role is played by DSOI, whereas the Landau level index is played by finite width induced energy subbands. In zigzag edge nanoribbon, increasing $\beta$ enhances the strength of inter-band  anticrossings. In particular, for $\beta/ta = 0.09$ [Fig.~\ref{zigzag_shc_band}], the spin-split subbands show visibly larger anticrossings, whereas for smaller $\beta$ the anticrossings remain relatively weak.

\section{Transport Properties of the Nanoribbon}

\label{sec:Transport Properties of the nanoribbon}

\subsection{Spin Hall conductance}

The SHC is calculated by using linear response based Kubo formula as used in Ref.~\cite{PhysRevLett126603}.  The SHC can now be written for a nanoribbon as  
%
\begin{equation}
\sigma^z_{xy} = \frac{e\hbar}{\Omega}  \sum_{\zeta \neq \zeta^\prime} 
(f_{\zeta } - f_{\zeta^\prime}) 
\frac{\text{Im} \left[ \langle \zeta | \hat{J}_x^z | 
\zeta^\prime \rangle \langle \zeta^\prime | \hat{v}_y | 
\zeta \rangle \right]}{(E_{\zeta^\prime} - E_{\zeta})^2},
 \label{equation shc}
\end{equation}
where  $\Omega=L_x\times L_y$ denotes the area, $f_{\zeta}=[1+\exp\{(E_\zeta-\mu)/(k_BT)\}]^{-1}$ is the Fermi-Dirac distribution function with $\mu$ being the chemical potential. 
Also, $\hat{J}_x^z$ is the spin current operator and is given by 
$\hbar \lbrace \sigma^z , \hat{v_x} \rbrace /4$, where the velocity operators 
are obtained 
$\hat{v}_y = (\hbar)^{-1} 
\partial H/\partial k_y$ and $\hat{v}_x= (i\hbar)^{-1}[H,x]$ 
with $x= \sum_n c_n^\dagger\, n\, c_n $~\cite{position_op}.
In the above Eq.~(\ref{equation shc}) we perform 
$\sum_{\zeta\ne\zeta^\prime}\rightarrow(L_y/2\pi)\int dk_y \sum_{n,n^\prime,s,s^\prime}$ 
to calculate the SHC for the nanoribbon geometry. 

\begin{figure}[h]
  \centering
  \includegraphics[height=6.6cm,keepaspectratio]{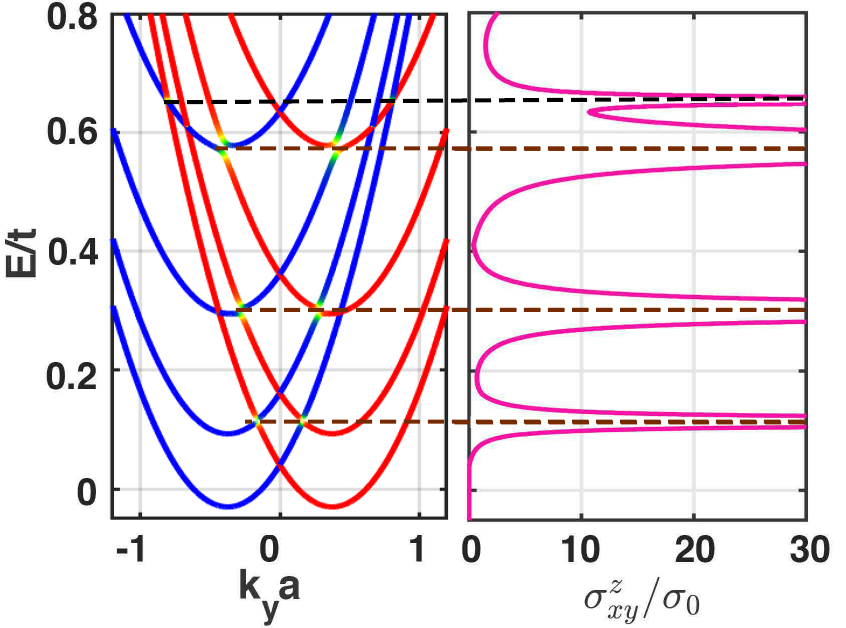}
  \caption{Spin resolved band structure (left) and SHC  (normalized by 
  $\sigma_0=e/8\pi$) are shown for a zigzag edge nanoribbon with width $N=21$ and 
  $\alpha = 0.1$, $\beta = 0.09$ in units of $ta$ at temperature $k_BT/t = 10^{-5}$. 
  The spin Hall resonances arising from anticrossings occur at chemical potentials $\mu/t=0.115,0.3,$ and $0.58$, highlighted by brown dashed lines.
The resonance associated with vanishing spin splitting appears at $\mu/t=0.65$, indicated by a black dashed line.}
  \label{fig:zigzag_SHR}
\end{figure}


We evaluate the SHC for zigzag edge nanoribbon and straight edge nanoribbon. The zigzag edge case is shown in the Fig.~\ref{fig:zigzag_SHR} whereas the straight edge is not shown here rather shifted to appendix \ref{app:straight_shc_band} as qualitative nature remain same. We observe that the SHC is enhanced sharply at a number of chemical potential, which correspond to the inter-band spin splitting vanishing or anticrossing points. This enhanced peaks are nothing but the so-called spin Hall resonance which precisely occur at the energies corresponding to the inter-band spin degeneracy points or anticrossing points in the band structure. Next, we also check that in the limiting cases of a pure Rashba system ($\beta = 0$) or a pure Dresselhaus system ($\alpha = 0$), the inter-band spin degeneracy points are strongly suppressed or absent, and as a result, the SHC does not exhibit any resonant peaks. This observation suggests that a relative competition between two types of spin splitting is required to achieve such resonance, here it is caused by RSOI and DSOI. We also mention that, the spin Hall resonances persist for both even and odd lattice terminations. Although the boundary condition modifies the subband dispersions, the qualitative structure of the resonances remains the same.

Our findings resonate well with the theoretical predictions made by Shen \textit{et al.}~\citep{PhysRevLett.92.256603}, where they studied resonant SHC in 2DEGs under a perpendicular magnetic field. In that work, the RSOI competes with the Zeeman splitting to induce energy level crossings that give rise to the resonance. In our case, an analogous mechanism takes place in nanoribbons via the interplay between RSOI and DSOI, resulting in anticrossings and degeneracy points between spin-resolved subbands and subsequent enhancement of the SHC.

%


\begin{figure}
    \centering
    \includegraphics[height=6.6cm,keepaspectratio]{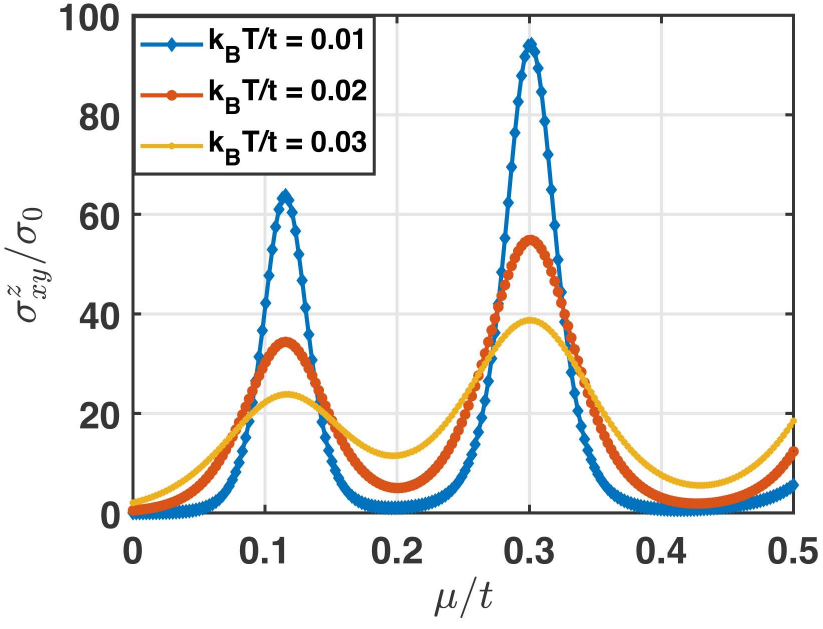}
    \caption{Spin Hall conductance versus chemical potential is plotted for zigzag edge 
    nanoribbon with width $N=21$ and $\alpha = 0.1$, $\beta = 0.09$ in units of $ta$ for 
    various temperatures.}
    \label{fig:shc_alltemp}
\end{figure}

We now comment on the physical origin of the different resonance widths visible in Fig.~\ref{fig:zigzag_SHR}. The width of the spin Hall resonances is mainly determined by thermal broadening and by the extent of mixing between the participating subbands. When the mixing between subbands of different spin character extends over a wider interval of $k_y$, the contributing states are distributed over a larger momentum range, resulting in an increased resonance width. In contrast, when the mixing is confined to a narrow region of $k_y$, the contributing states are limited and the corresponding resonance remains sharp. This behavior is directly reflected in the Fig.~\ref{fig:zigzag_SHR}-~\ref{fig:straight_SHR}, where broader resonances are observed at crossings and anticrossings characterized by an extended momentum range of spin mixing.

It is noteworthy to mention that when $\alpha = \beta$, the system exhibits a high degree of symmetry, 
and although many spin degeneracy points are visible in the band structure of both zigzag and straight edge as shown in Fig.~\ref{zigzag_band_equal}-\ref{straight_band_equal}, the matrix elements of the spin current operator vanishes. Consequently, the SHC is identically zero across the entire range of chemical potentials. Also in such case the Hamiltonian commutes with the spin operator $\sigma_x-\sigma_y$~\cite{PhysRevLett.97.236601}, leading to the conservation of this spin component, this conserved spin symmetry forbids any net spin Hall response. This vanishing of SHC for $\alpha=\beta$ is not a finite size effect of nanoribbon but persists in the bulk system also~\cite{PhysRevB.71.085315}. We also comment that, as the spin Hall phenomenon occurs in a $2$D fermionic system, it is desirable to keep width also quite large compared to the present case. Here, we keep width very narrow as $N=21$ only to make degeneracy or anticrossing points clearly visible. With the further increase of width a large number of subbands will be formed and many such degeneracy or anticrossing points exhibiting resonance in SHC will occur. 

Next, we briefly discuss the spin Hall phenomenon in a $2$D  electronic system with an 
anisotropic RSOI and DSOI. The ab-initio calculation based theoretical prediction in the surface states of $Au(110)$ suggested the existence of an anisotropic Rashba splitting~\cite{Nagano_2009}. In fact the $C_{2\nu}$ point-group symmetry analysis not only suggested the mass anisotropy but also hinted an anisotropic RSOI~\cite{PhysRevB.78.195412} as $H_R=\alpha_{x}k_x\sigma_y+\alpha_y k_y\sigma_y$ where $\{\alpha_x,\alpha_y\}$ are two independent parameters describing the degree of spin splitting along two orthogonal directions in momentum space. Similarly, anisotropic Dresselhaus interaction can be modeled as $H_D=\beta_{x}k_x\sigma_x-\beta_y k_y\sigma_y$ with $\{\beta_x,\beta_y\}$ denoting the anisotropy in the DSOI. In the discrete lattice model, such anisotropy can be captured by considering a weak imbalance between hopping parameters along two orthogonal directions as $t_x$ and $t_y$, as $t_y=\nu t_x$ where $\nu$ describes the degree of anisotropy. Following the same approach as in the isotropic Rashba–Dresselhaus case,  we note that the spin degeneracy points and the anticrossing points persist even in the presence of anisotropy. The SHC continues to exhibit resonance whenever the chemical potential lies at these points. It indicates that inter-subband resonance phenomenon in SHC is robust to weak anisotropy in band structure. 

We now turn to the role of finite temperature on the SHC, we find that the suppression of SHC at finite temperature is more pronounced at the lower energy degeneracy points compared to the higher energy ones Fig.~\ref{fig:shc_alltemp}. This behavior can be understood in terms of the difference between the Fermi occupation factors. When the chemical potential lies in the lower bands, the upper band remains nearly unoccupied and hence the difference is essentially governed by the lower band Fermi function. In this regime, the thermal broadening of the Fermi function strongly reduces the SHC resonance. However, when the chemical potential is located in the higher bands, both the upper and lower bands are partially occupied. The effect of broadening then tends to cancel out between the two bands, leading to a much weaker suppression of SHC at the higher energy degeneracy and anticrossing points. Hence, we conclude that the resonance features in the higher bands are more robust to temperature effects.

We comment here that the typical sample that could be suitable for resonance phenomenon is InAs quantum wells in which both RSOI and DSOI with almost the same order of strength has been already confirmed by photocurrent measurement~\cite{PhysRevLett.92.256601}. What we require is to reduce the width to engineer a number of subbands. To achieve the resonance in SHE, it is important to perform the fine tuning of chemical potential by gate voltage so that it capture the inter subband spin degenerate or anticrossing points. The experimentally relevant nanoribbon widths typically lie in the range of $50–150nm$~\cite{PhysRevB.94.035444}. For InAs square lattice discretization with lattice constant $a=0.6nm$, a ribbon width of $W=77nm$ corresponds to $N \approx 128$ lattice sites across the transverse direction, placing our simulations well within experimentally accessible regimes. The strength of RSOI in InAs is known to be relatively large and electrically tunable, with experimentally reported values in the range $\alpha=(0.5-3)\times 10^{-11} eVm$, and can be tuned by gating. Using the InAs effective mass $m^*=0.02m_e$~\cite{PhysRev.105.460} and lattice spacing of InAs, experimentally reported Rashba couplings correspond to a dimensionless lattice Rashba strength $\alpha/t=10^{-3}-10^{-5}$. Importantly, while a single global gate generally changes both the chemical potential $\mu$ and the interfacial electric field and hence $\alpha$ experiments employing dual gate geometries have demonstrated that these two effects can be largely decoupled~\cite{PhysRevB.94.035444,D2NA00143H,doi:10.1021/nl301325h}. In realistic device geometries, $\alpha$ can be varied independently of $\mu$, justifying our theoretical approach of treating the chemical potential and Rashba coupling as independently tunable parameters.

\subsection{Spin Nernst coefficient}
Having evaluated the SHC of the nanoribbon, we now extend our analysis to its thermal analogue by computing the spin Nernst coefficient. The SNC characterizes the transverse spin current generated in response to a longitudinal temperature gradient, and is defined as
\(\alpha^{z}_{xy} = - J^{z}_{x} / \nabla_{y} T\). Within linear response theory, SNC is obtained as~\cite{PhysRevB.96.224415,PhysRevB.110.174411,PhysRevB.88.201108}

\begin{equation}
\alpha^{z}_{xy}
= \frac{1}{T}
\sum_{\zeta} 
\Omega^{z,\zeta}_{xy}
\Lambda_{\zeta}
\label{eq:alphaSN}
\end{equation}
with
\begin{equation}
\begin{aligned}
\Omega^{z,\zeta}_{xy}
&=
-2\,\mathrm{Im}
\sum_{\zeta' \neq \zeta}
\frac{
\langle \zeta | \hat{J}^{z}_{x} | \zeta' \rangle
\langle \zeta' | \hat{v}_{y} | \zeta \rangle
}{
\left( E_{\zeta} - E_{\zeta'} \right)^{2}
},
\\[-2pt]
\Lambda_{\zeta}
&=
(E_{\zeta}-\mu)\,f_{\zeta}
+ k_{B}T\,\ln\!\bigl(1 + e^{-\beta(E_{\zeta}-\mu)}\bigr)
\end{aligned}
\end{equation}
\begin{figure}
\includegraphics[scale=0.6]{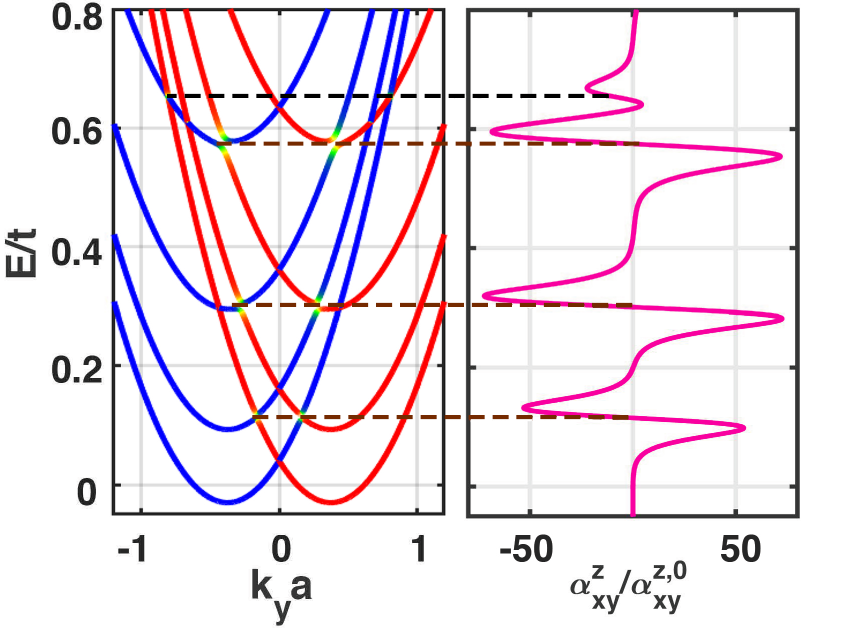}
\caption{spin resolved band structure (left) and spin Nernst coefficient normalized with 
  $\alpha_{xy}^{z,0}=-k_B/(4\pi)$ for a zigzag edge nanoribbon with width $N=21$ and 
  $\alpha = 0.1$, $\beta = 0.09$ in units of $ta$ at temperature $k_BT/t = 0.01$.}
  \label{fig:zigzag_SNC}
\end{figure}
 
The computed SNC as a function of the chemical potential, together with the corresponding band structure, is shown in Fig.~\ref{fig:zigzag_SNC}. We observe a sequence of sharp positive and negative peaks that occur precisely when the chemical potential crosses the anticrossing or spin degeneracy points in the spectrum. These features are fully consistent with the behaviour of the SHC, which exhibits spin Hall resonance at the same points.
\begin{equation}
\alpha^{z}_{xy}
= -\,\frac{\pi^{2}}{3e}\,k_{B}^{2}T\,\sigma^{z}_{xy}(\mu)'
\label{mott}
\end{equation}

This behavior is also consistent with the thermoelectric Mott relation, Eq.~(\ref{mott}) ~\cite{PhysRevB.96.224415}, which links the SNC to the energy derivative of the SHC. Thus, whenever the SHC undergoes a rapid change such as at anticrossings or near spin degenerate points, its derivative becomes large, giving rise to the pronounced peaks in the SNC. The overall structure of the SNC therefore provides a clear thermoelectric fingerprint of the underlying anticrossing and spin degenerate points.

Next, we briefly check whether the longitudinal transport exhibits any signatures of vanishing spin splitting or anticrossings.
\subsection{Longitudinal transport}  

The longitudinal conductivity characterizes the charge transport along the translationally invariant direction under the application of an external bias. In order to capture the contribution of finite width induced subbands, it is convenient to use the real space Green function based approach developed by Sancho  \textit{et al.}~\cite{sancho1984,sancho1985}.  In this approach the longitudinal direction is taken to be very long compared to the width, and both the ends of the ribbon are attached to two leads of identical system. Following this approach 
it is straight forward to write the transmission between two leads though central region at $E$ as~\cite{C_Caroli_1971}
\begin{equation}
\mathcal{T}(E)=\mathrm{Tr}\!\left[\Gamma_L\, G_{11}(E)\, \Gamma_R\, G_{11}^\dagger(E)\right]
\label{eq:TotTrace}.
\end{equation}
where  
\begin{equation}
G_{11} = \left[ (E + i\eta)I - H_{11} - \Sigma_L - \Sigma_R \right]^{-1}
\end{equation}  
is the retarded Green’s function of the scattering region including the effects of the leads attached to both ends. The leads are encoded into the above total Green function through $2N\times 2N$ matrices \(\Gamma_{L,R}=i(\Sigma_{L,R}-\Sigma_{L,R}^\dagger)\) where $\Sigma_L$ and $\Sigma_R$ are self energy function due to left and right lead, respectively.


At zero temperature, the longitudinal conductance follows from the 
Landauer–Büttiker relation~\citep{Landauer1970,PhysRevLett.57.1761},
\begin{equation}
G(E)=\frac{e^2}{h}\mathcal{T}(E),
\end{equation}

\begin{figure}
\includegraphics[height=6.6cm, keepaspectratio]{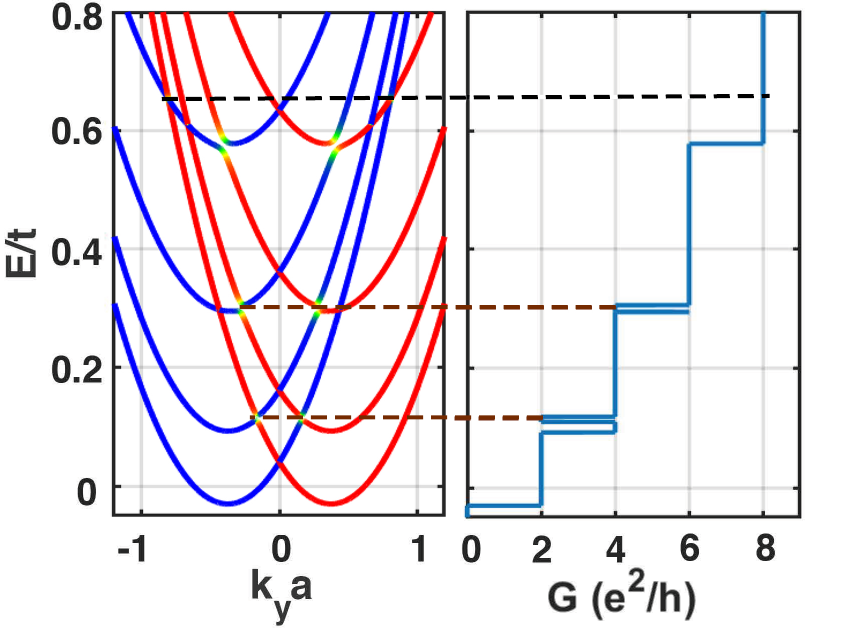}
\caption{Band structure (left) and corresponding longitudinal conductance (right, in units of $e^2/h$ )
 for a zigzag edge nanoribbon with $\alpha = 0.1$, $\beta = 0.09$ in units of $ta$ and the width is $ N = 21$. 
The brown dashed lines highlight interband anticrossings while the black dashed lines show interband spin degeneracy point.}
\label{zigzag_longitudinal_c}
\end{figure}

By using Eq.~(\ref{eq:TotTrace}) we compute zero temperature conductance as a function of chemical potential against the corresponding band structure for zigzag edge nanoribbon in the Fig.~\ref{zigzag_longitudinal_c}.  As usual, the conductance exhibits quantization in units of $2e^2/h$ when $\mu <0.1$ . This can be attributed to the fact that only one subband with two electron incident channels is occupied by the chemical potential. As the chemical potential starts occupying the spin up subband of the lowest subband, both bands contribute one incident channel each maintaining the conductance again at $2e^2/h$. At $\mu \approx 0.1$ the conductance suddenly increases to $4e^2/h$ reflecting the existence of four electron incident channels, three from the spin up subband and one from the spin down subband. However, within very  a very narrow range of $\mu$ around $0.1$, the conductance drops back to $2e^2/h$ and then rises up again to $4e^2/h$. This unusual feature originates from the presence of an anticrossing point at $\mu=0.1$.  With further increase in $\mu$, the phenomenon reappears at $\mu=0.3$ corresponding to another anticrossing point. Such phenomenon occurs because of the drops of number of electron incident channel from the subband occupied by chemical potential. We emphasize that such behavior does not occur at the inter-subband spin degeneracy points. Hence we can conclude that although a signature of anticrossing point is visible in longitudinal conductance, it shows no corresponding features at the inter-band spin degeneracy. Finally, we note that the qualitative features of the conductance remain the same for the straight edge case, with the only difference being a shift in the locations of the anticrossing points.  

\section{Summary}
\label{sec:summary}
We have investigated the spin-Hall transport and longitudinal transport for a 2D electronic system with RSOI and DSOI in a discrete lattice model. We have considered square lattice system with straight edge and zigzag edge geometry. In both cases, we showed that a number of spin degeneracy points and anticrossing points between two subbands with opposite spin emerges. The SHC shows a sharp enhancement when the chemical potential is situated at those spin degeneracy points or anticrossing points, exhibiting spin-Hall resonance. Hence, we conclude that resonance in SHC can be achieved without applying any external perturbation like magnetic field or light field as proposed previously. However such resonance can occur only if RSOI and DSOI both are present. We have also investigated the effects of anisotropy and finite temperature and demonstrated that the latter results in suppression of SHC but this suppression is weaker in higher energy degeneracy and anticrossing points while the former results in a shift of the position of resonance points in the SHC. We also study the spin Nernst effect and find that the spin Nernst coefficient reflects these anticrossing and spin degeneracy points through clear thermoelectric signatures. Finally we study the longitudinal conductance by using retarded Green function approach, and reveal that anticrossing point can give a signature to the conductance whereas the degeneracy point does not leave any mark. 

\section{Acknowledgement}
SK Firoz Islam acknowledges financial support by the project ANRF/ECRG/2024/005166/PMS.

\appendix
\section{Straight edge nanoribbon: Band structure and Transport properties}
\label{app:straight_shc_band}

\begin{figure}
\includegraphics[height=6.6cm, keepaspectratio]{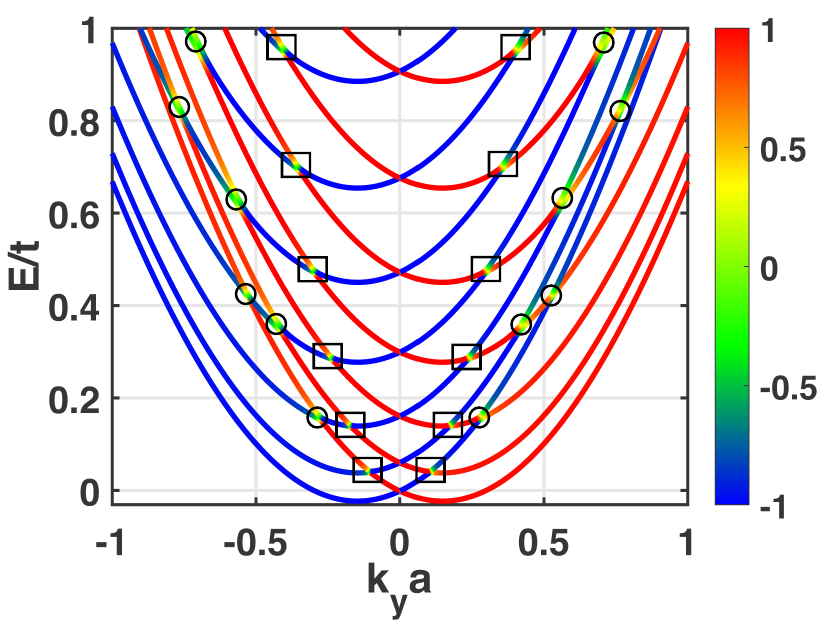} 
\caption{spin resolved band dispersion of a straight edge nanoribbon for
$\alpha=0.11$ and $\beta=0.1$ in units of $ta$.
The color scale represents the expectation value of the in-plane spin operator,
i.e., the spin polarization in units of $\hbar/2$.
Anticrossing points are highlighted by squares, while spin degeneracy points are
marked by circles.}
\label{straight_shc_band}
\end{figure}

\begin{figure}
\includegraphics[height=6.6cm, keepaspectratio]{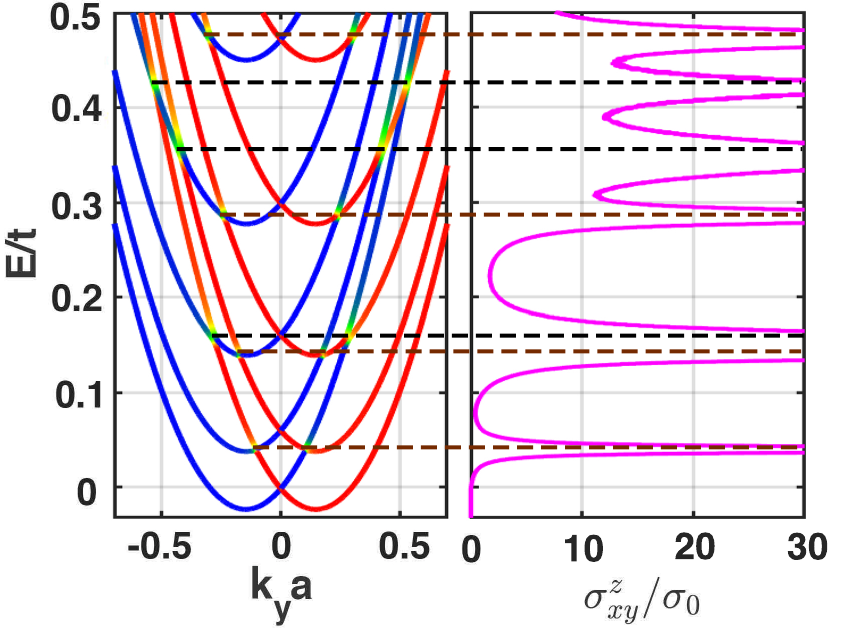}
\caption{Spin resolved band structure (left) and SHC (in units of  $\sigma_0=e/8\pi$) for a straight edge nanoribbon are plotted for width $N=21$,
  $\alpha= 0.11$, $\beta = 0.1$ in units of $ta$, at temperature $k_BT/t = 10^{-5}$. 
  The spin Hall resonances arising from anticrossings occur at chemical potentials $\mu/t=0.0418,0.14105,0.28511,$ and $0.47312$, highlighted by brown dashed lines.
The resonance associated with vanishing spin splitting appears at $\mu/t=0.15919,0.359$ and $0.42526$, indicated by a black dashed line.}
\label{fig:straight_SHR}
\end{figure}

We briefly discuss the case of a straight edge nanoribbon here. We present lowest few spin-split subbands for straight edge nanoribbon in Fig.~\ref{straight_shc_band}  where red and blue curves represent the spin-up and spin-down components, respectively. Very much similar to the case of zigzag ribbon, here also a number of spin splitting vanishing as well as anticrossing points are present between two subbands, which are highlighted by circles and squares respectively. It is noticeable that the number of such points are relatively higher in straight edge case compared to zigzag edge. It is expected that the SHR can occur when the chemical potential passes through the inter subband spin degenerate or anticrossing points. The SHC is plotted in the Fig.~\ref{fig:straight_SHR} showing multiple resonances. 

\begin{figure}
\includegraphics[height=6.6cm, keepaspectratio]{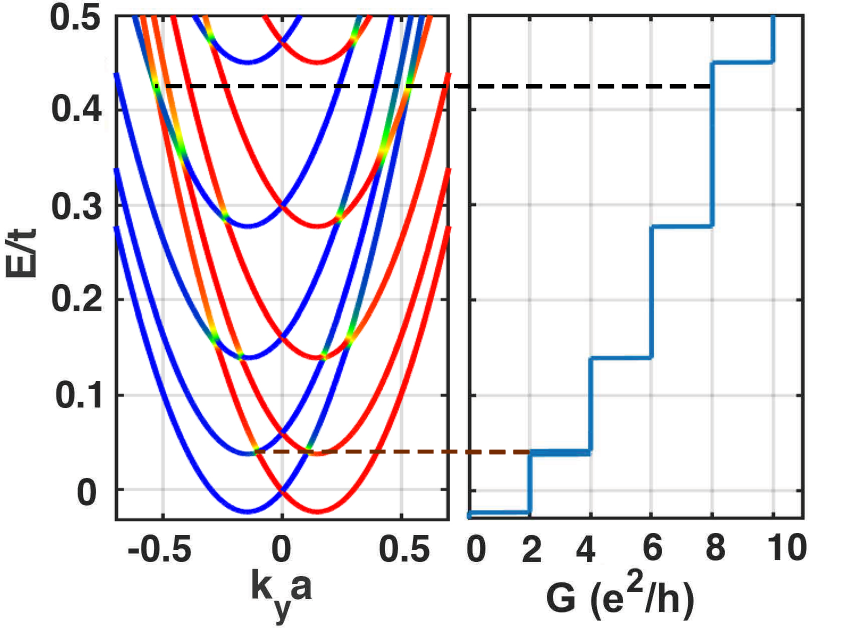}
\caption{spin resolved band structure (left) and corresponding longitudinal conductance (right, in units of $e^2/h$ )
 for a straight edge nanoribbon with $\alpha = 0.11$, $\beta = 0.1$ in units of $ta$ and the width is $ N = 21$. 
The brown dashed lines highlight interband anticrossings while the black dashed lines show interband spin degeneracy point.}
\label{straight_longitudinal_c}
\end{figure}

The longitudinal conductance for straight edge ribbon is also plotted in Fig.~\ref{straight_longitudinal_c} which also shows quantized steps in units of $2e^{2}/h$, with the overall features similar to the zigzag case. 



\section{Full Band Structure of Zigzag- and Straight Edge Nanoribbons}
\label{equal_alphabeta}
\begin{figure}
\includegraphics[height=6.6cm,keepaspectratio]{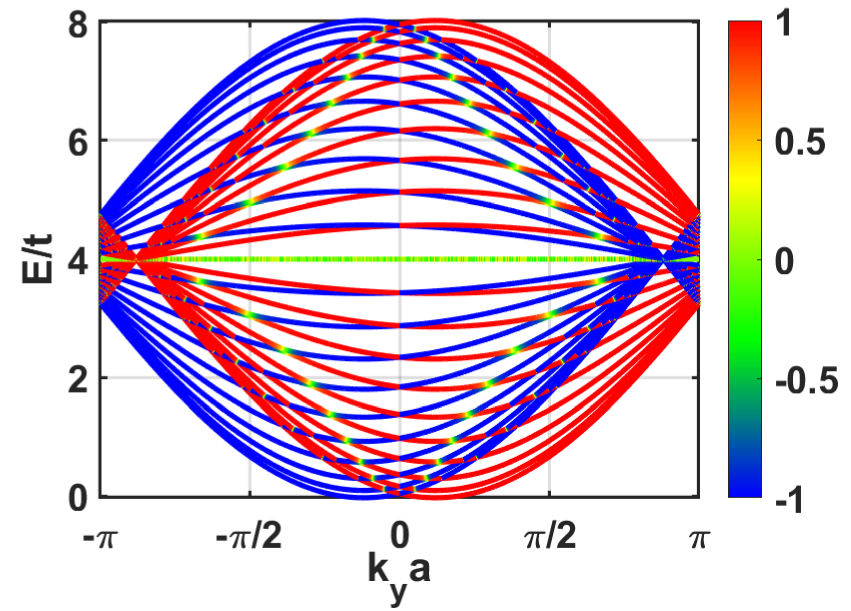} 
\caption{Spin resolved band dispersion of a zigzag edge nanoribbon is presented for
$\alpha=0.1$ and $\beta=0.09$ in units of $ta$.
The color scale represents the expectation value of the in-plane spin operator,
i.e., the spin polarization in units of $\hbar/2$.}
\label{zigzag_band_full}
\end{figure}

\begin{figure}
\includegraphics[height=6.6cm,keepaspectratio]{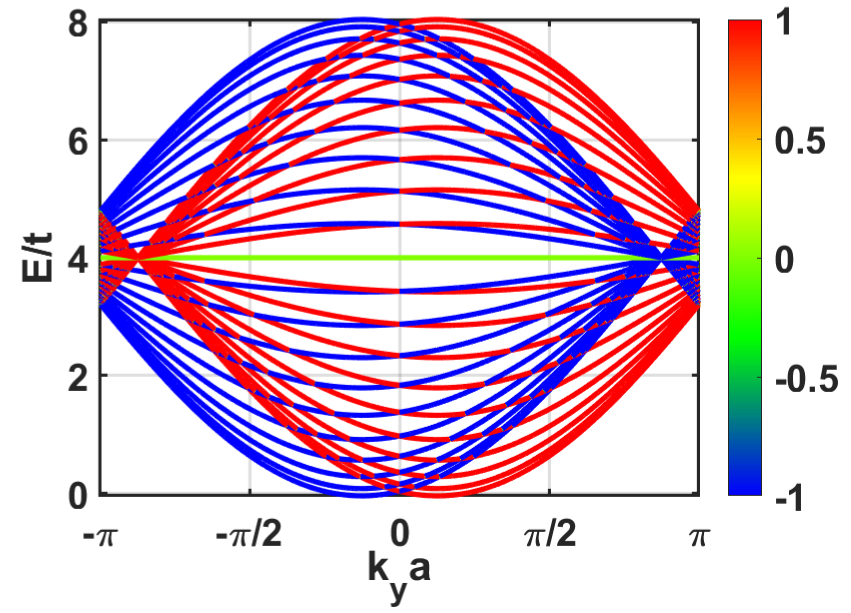} 
\caption{Spin resolved band dispersion of a zigzag edge nanoribbon is plotted for
$\alpha=0.1$ and $\beta=0.1$ in units of $ta$.
The color scale represents the expectation value of the in-plane spin operator,
i.e., the spin polarization in units of $\hbar/2$.} 
\label{zigzag_band_equal}
\end{figure}

\begin{figure}
\includegraphics[height=6.6cm, keepaspectratio]{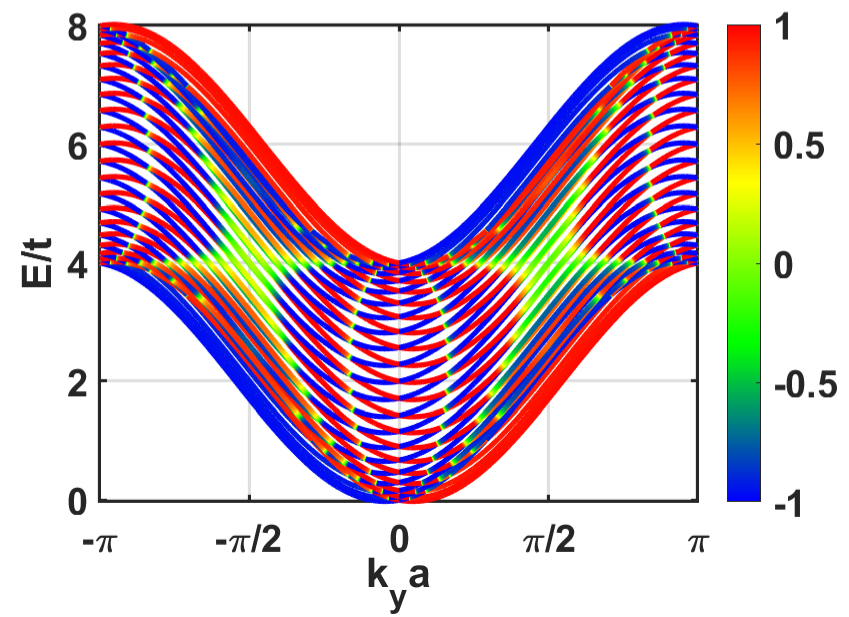} 
\caption{Spin resolved band dispersion of a straight edge nanoribbon is plotted for
$\alpha=0.11$ and $\beta=0.1$ in units of $ta$.
The color scale represents the expectation value of the in-plane spin operator,
i.e., the spin polarization in units of $\hbar/2$.}
\label{straight_band_full}
\end{figure}

In this appendix, we present the full spin resolved band structures of both zigzag edge and straight edge nanoribbons to provide an overall view of the energy spectrum and to illustrate the role of edge geometry. Figures~\ref{zigzag_band_full}–\ref{straight_band_full} display the complete band dispersions across the Brillouin zone.
For the zigzag edge nanoribbon, edge localized states appear at $E=4t$, while such states are absent in the straight edge geometry. This distinction is consistent with earlier results reported for systems with pure RSOI in Ref.~\cite{hubbardrashba}, and we find that this qualitative feature remains unchanged in the simultaneous presence of RSOI and DSOI. Since these edge localized states occur at relatively high energies and accessing such large chemical potentials is not physical for typical semiconductor nanostructures, the main analysis in the paper is restricted to the lowest few subbands.

We further clarify the qualitative difference between the full band structures of straight edge and zigzag edge nanoribbons. Although the lowest few subbands in both geometries are qualitatively similar and exhibit conventional parabolic dispersion, a marked difference emerges at higher energies. In the zigzag edge nanoribbon,  half of the subbands become inverted, above the flat edge band. Such band inversion is absent in the straight edge nanoribbon, where all subbands retain non-inverted, parabola like dispersion throughout the spectrum. As a consequence, within a given energy interval, the straight edge nanoribbon accommodates a larger number of subbands compared to the zigzag edge case. This increased subband density naturally leads to a higher number of band crossings and anticrossings, which is clearly reflected in Fig.~\ref{straight_shc_band}.

When $\alpha \neq \beta$, pronounced anticrossings and spin degeneracy points emerge between spin split subbands accompanied by strong spin mixing, as indicated by the continuous variation of the spin polarization near the crossing and anticrossing points. In contrast, for $\alpha=\beta$, the band structure exhibits numerous spin degeneracy points distributed throughout the Brillouin zone. In this case, anticrossings are absent and only spin degenerate crossings appear. The degeneracies at any finite $k_y$ are of the same order as those at  $k_y=0$ and persists for both zigzag and straight edge nanoribbon geometries. Moreover we see that when $\alpha=\beta$ the mixed spin states are absent near crossing points. As a consequence, the interband matrix elements responsible for SHC are suppressed, leading to a vanishing spin Hall response despite the presence of multiple spin degeneracy points in the band structure.


\begin{figure}
\includegraphics[height=6.6cm, keepaspectratio]{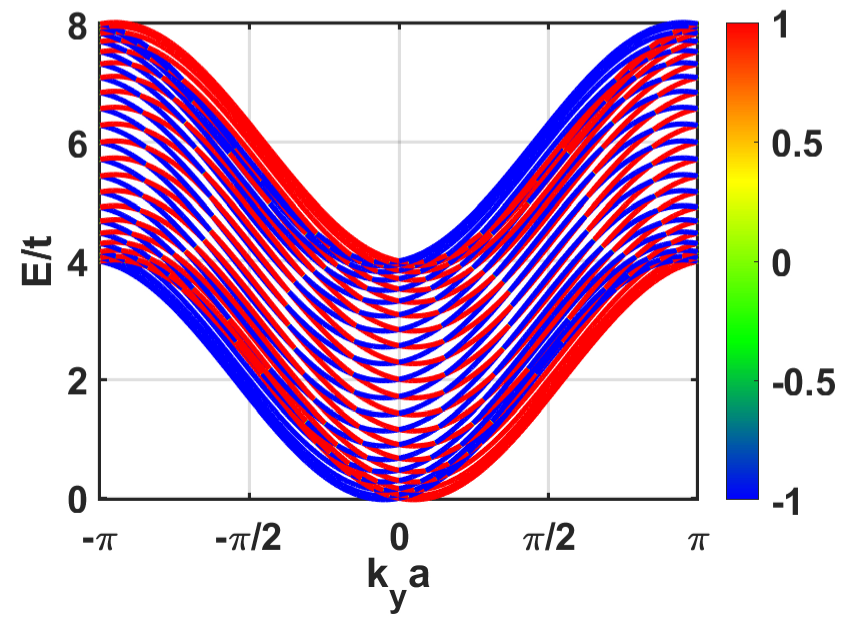} 
\caption{Spin resolved band dispersion of a straight edge nanoribbon is plotted for
$\alpha=0.11$ and $\beta=0.11$ in units of $ta$.
The color scale represents the expectation value of the in-plane spin operator,
i.e., the spin polarization in units of $\hbar/2$.}
\label{straight_band_equal}
\end{figure}
\clearpage
\bibliography{manuscript_3_1}
\end{document}